\documentclass[a4paper,square,sort,comma,numbers,11pt]{article}
\usepackage{jheppub} 

\usepackage{upgreek} 
\usepackage{float}
\usepackage{graphicx}   
\usepackage{amsmath}
\usepackage{ulem}
\usepackage{lineno}

\title{\boldmath Searching for millicharged particles with 1~kg of Skipper-CCDs using the NuMI beam at Fermilab}

\author[1,2]{Santiago Perez}
\author[1,2]{Dario Rodrigues}
\author[3]{Juan Estrada}
\author[3]{Roni Harnik}
\author[4]{Zhen Liu}
\author[5]{Brenda A. Cervantes-Vergara}
\author[5]{Juan Carlos D'Olivo}
\author[6]{Ryan D. Plestid}
\author[3]{Javier Tiffenberg}
\author[7]{Tien-Tien Yu}
\author[5]{Alexis Aguilar-Arevalo}
\author[8]{Fabricio Alcalde-Bessia}
\author[8]{Nicol\'as Avalos}
\author[9]{Oscar Baez}
\author[3]{Daniel Baxter}
\author[8]{Xavier Bertou}
\author[10]{Carla Bonifazi}
\author[3]{Ana Botti}
\author[3]{Gustavo Cancelo}
\author[11]{Nuria Castelló-Mor}
\author[12]{Alvaro E. Chavarria}
\author[3, 13]{Claudio R. Chavez}
\author[13]{Fernando Chierchie}
\author[9]{Juan Manuel De Egea}
\author[14]{Cyrus Dreyer}
\author[3, 15]{Alex Drlica-Wagner}
\author[14]{Rouven Essig}
\author[8]{Ezequiel Estrada}
\author[16]{Erez Etzion}
\author[17]{Paul Grylls}
\author[3]{Guillermo Fernandez-Moroni}
\author[14]{Marivi Fern\'andez-Serra}
\author[9]{Santiago Ferreyra}
\author[18]{Stephen Holland} 
\author[11]{Agustín Lantero Barreda}
\author[3]{Andrew Lathrop}
\author[17]{Ian Lawson}
\author[19]{Ben Loer}
\author[17]{Steffon Luoma}
\author[3,15]{Edgar Marrufo Villalpando}
\author[5]{Mauricio Martinez Montero}
\author[12]{Kellie McGuire}
\author[9]{Jorge Molina}
\author[15]{Sravan Munagavalasa}
\author[15]{Danielle Norcini}
\author[12]{Alexander Piers}
\author[15]{Paolo Privitera}
\author[3]{Nathan Saffold}
\author[19]{Richard Saldanha}
\author[14]{Aman Singal}
\author[15]{Radomir Smida}
\author[20]{Miguel Sofo-Haro}
\author[9]{Diego Stalder}
\author[3]{Leandro Stefanazzi}
\author[12]{Michelangelo Traina}
\author[3]{Yu-Dai Tsai}
\author[3]{Sho Uemura}
\author[21]{Pedro Ventura}
\author[11]{Rocío Vilar Cortabitarte}
\author[15]{Rachana Yajur}

\affiliation[1]{\normalsize\it Universidad de Buenos Aires, Facultad de Ciencias Exactas y Naturales, Departamento de Física. Buenos Aires, Argentina.}
\affiliation[2]{\normalsize\it CONICET - Universidad de Buenos Aires, Instituto de Física de Buenos Aires (IFIBA). Buenos Aires, Argentina}
\affiliation[3]{Fermi National Accelerator Laboratory, IL, USA}
\affiliation[4]{School of Physics and Astronomy, University of Minnesota, Minneapolis, Minnesota, 55455 USA}
\affiliation[5]{Universidad Nacional Aut\'onoma de M\'exico, Ciudad de M\'exico, M\'exico}
\affiliation[6]{
Walter  Burke Institute for Theoretical Physics, California Institute of Technology, Pasadena, CA 91125}
\affiliation[7]{Department of Physics and Institute for Fundamental Science, University of Oregon, Eugene, Oregon 97403, USA}
\affiliation[8]{Centro Atomico Bariloche, Rio Negro, Argentina}
\affiliation[9]{Facultad de Ingenier\'ia, Universidad Nacional de Asunci\'on, Paraguay}
\affiliation[10]{International Center of Advanced Studies and Instituto de Ciencias Físicas, ECyT-UNSAM and CONICET, Argentina}
\affiliation[11]{Instituto de Fisica de Cantabria, Santander, Spain }
\affiliation[12]{University of Washington, WA, USA}
\affiliation[13]{IIIE CONICET and DIEC Universidad Nacional del Sur, Argentina}
\affiliation[14]{Stony Brook University, NY, USA}
\affiliation[15]{University of Chicago, IL, USA}
\affiliation[16]{Tel Aviv University, Israel}
\affiliation[17]{SNOLAB, ON, Canada}
\affiliation[18]{Lawrence Berkeley National Laboratory, CA, USA}
\affiliation[19]{Pacific Northwest National Laboratory, WA, USA}
\affiliation[20]{Universidad Nacional de C\'ordoba, Instituto de F\'isica Enrique Gaviola (CONICET) and Reactor Nuclear RA0 (CNEA), C\'ordoba, Argentina.}
\affiliation[21]{Instituto de F\'isica, Universidade Federal do Rio de Janeiro, Rio de Janeiro, RJ, Brazil}

\emailAdd{santiep.137@gmail.com, rodriguesfm@df.uba.ar}

\abstract{Oscura is a planned light-dark matter search experiment using Skipper-CCDs with a total active mass of 10~kg. As part of the detector development, the collaboration plans to build the Oscura Integration Test (OIT), an engineering test with 10\% of the total mass.
Here we discuss the early science opportunities with the OIT to search for millicharged particles (mCPs) using the NuMI beam at Fermilab. 
mCPs would be produced at low energies through photon-mediated processes from decays of scalar, pseudoscalar, and vector mesons, or direct Drell-Yan productions. Estimates show that the OIT would be a world-leading probe for mCPs in the $\sim$MeV mass range.}

\begin{document}
\maketitle
\flushbottom

\section{Oscura Experiment}
\label{sec:intro}
\subsection{Science goals}

Skipper-CCDs with ultra-low noise are among the most promising detector technologies for the construction of a multi-kg experiment probing electron recoils from sub-GeV dark matter (DM). In 2017, the ability to precisely measure the number of free electrons in each of the million pixels across a Skipper-CCD was demonstrated ~\cite{Tiffenberg:2017aac} using sensors designed by the Lawrence Berkeley National Laboratory (LBNL) Micro Systems Lab. As part of an ongoing light DM search program, SENSEI ("Sub-Electron Noise Skipper-CCD Experimental Instrument") took data with small scale ($\sim$1g and $\sim$2g) Skipper-CCD detectors at Fermilab on the surface and underground, setting world-leading constraints on DM-electron interactions for DM masses in the range of 500~keV to 5~MeV~\cite{sensei2018,sensei2019,SENSEI:2020dpa}. The SENSEI Collaboration is now commissioning a 100 g experiment with the same technology at SNOLAB. Furthermore, the DAMIC-M Collaboration is planning to use Skipper-CCDs for a 1 kg experiment in the coming years \cite{2020DAMICM}.  
The largest planned experiment using this technology is called Oscura, which will implement 10 kg of Skipper-CCDs for the search of light dark matter.

\subsection{Experiment Design}

\begin{figure}[t] 
    \centering
	\includegraphics[width=0.3\linewidth]{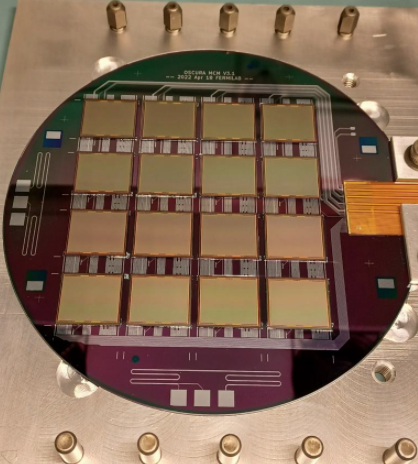}  	\includegraphics[width=0.65\linewidth]{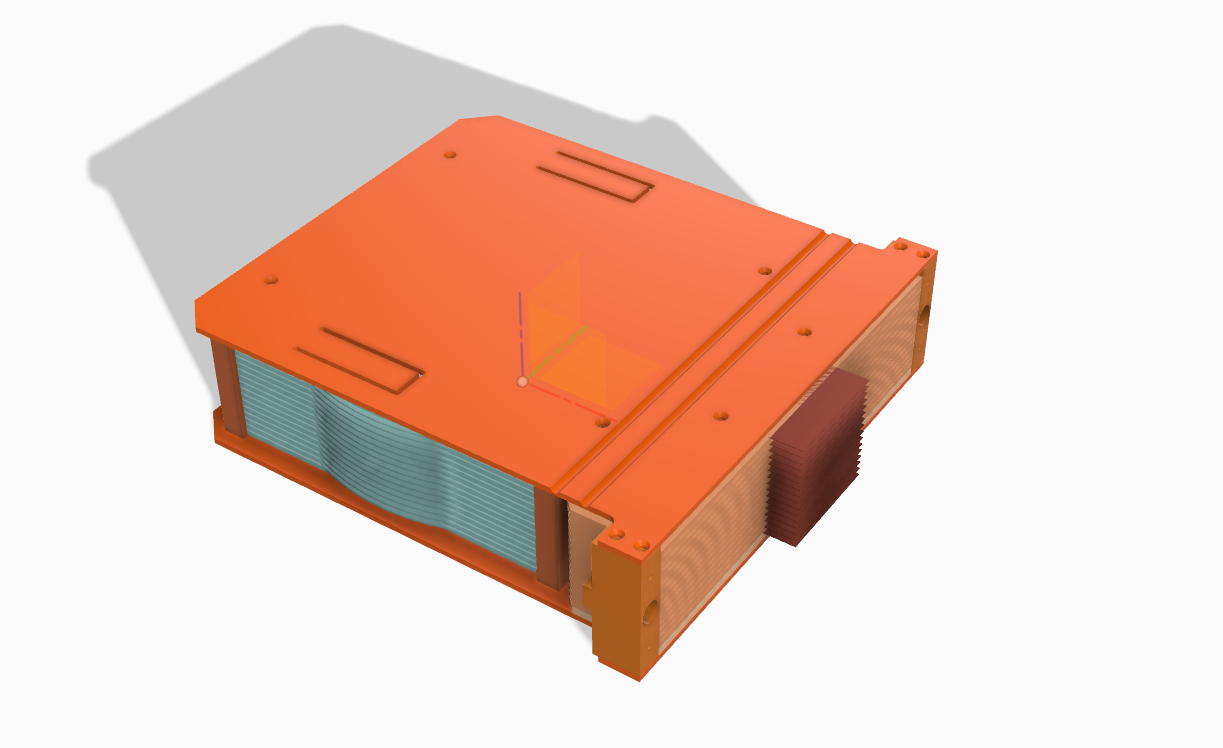}
	\caption{(Left) Fully assembled MCM with 16 Oscura prototype sensors. (Right) Oscura Super Module (SM) with 16 MCMs supported and shielded with electro-formed copper.}
	\label{fig:MCM}
\end{figure}

\begin{figure}[t!]
    \centering
    \includegraphics[width=0.3\textwidth]{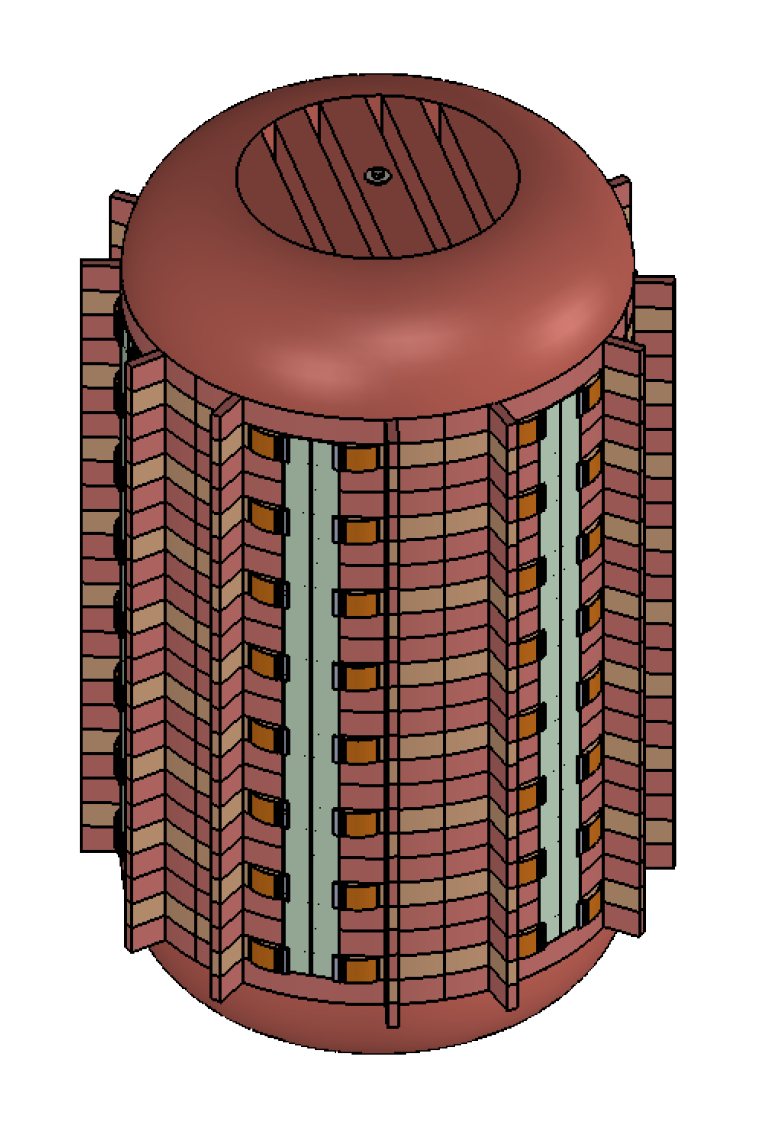} 
    \caption{Full detector payload and internal shield.}
    \label{fig:fullpayload}
\end{figure}

\begin{figure}[ht]
    \centering
    \includegraphics[width=0.5\textwidth]{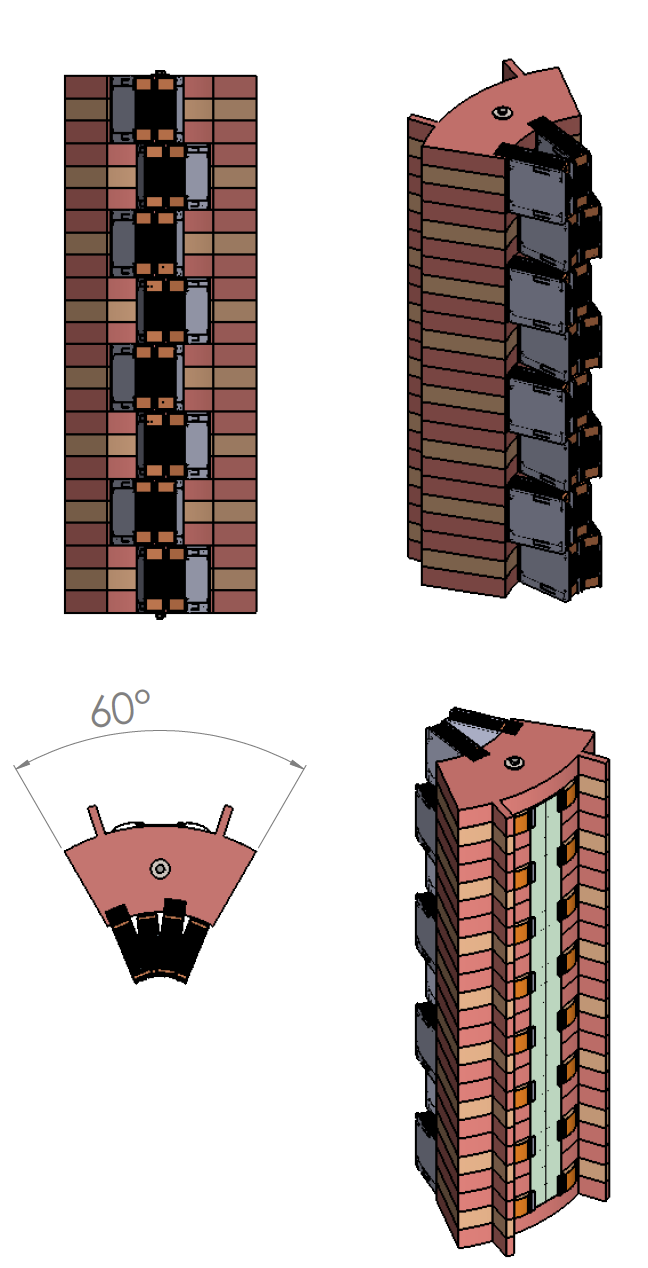} 
    \caption{Different views of an Oscura slice and the lead shield (pink). Oscura Integration Test will consist of one slice of the Oscura experiment with 6 inches of shield in all directions. Each slice has 16 supermodules arranged in 8 pairs. The full experiment will be composed of six slices forming the payload cylinder shown in Fig.~\ref{fig:fullpayload}.
    }
    \label{fig:slice}
\end{figure}

The Oscura design is based on 1.35 Mpix sensors~\cite{oscuraSensors} packaged on a Multi-Chip-Module (MCM)~\cite{Oscura2022}. Each MCM consists of 16 sensors mounted on a 150 mm diameter silicon wafer with traces connecting the CCD to a low radiation background flex circuit. The package is designed to keep only low-background materials next to the active volume of the CCD. Figure \ref{fig:MCM} shows pictures of an MCM fully populated with CCDs. The MCMs will be arranged into Super Modules (SM), and each SM will hold 16 MCMs using a support structure of ultrapure electro-deposited copper \cite{electroformedcopper}. The SM also includes copper to shield the radiation of the first couple of centimeters of the flex circuit from the sensors. The Oscura experiment needs 79 SMs to reach 10~kg active mass. Figure~\ref{fig:MCM} shows the design of a SM.
The full detector payload with 96 SMs inside the lead shield is shown in Fig.~\ref{fig:fullpayload}, and the detector payload is arranged as a cylinder formed by six columnar slices (see Fig.~\ref{fig:slice}).

The operation of Skipper-CCDs with low dark current requires cooling down the system to between 120~K and 140~K (the optimal operating point will be determined from the prototype sensors). The current strategy for the cooling system is to submerge the full detector array in a Liquid Nitrogen (LN$_2$) bath operated with a vapor pressure of 450 psi to reach this temperature. We are also considering the alternative design of using nitrogen gas to cool down the detectors uniformly, thereby reducing the pressure requirements of the vessel. Nitrogen gas has less thermal conductivity than LN$_2$, but preliminary simulations indicate that this simpler solution is feasible. 

\begin{figure}[!ht]
    \centering
    \includegraphics[width=0.5\textwidth]{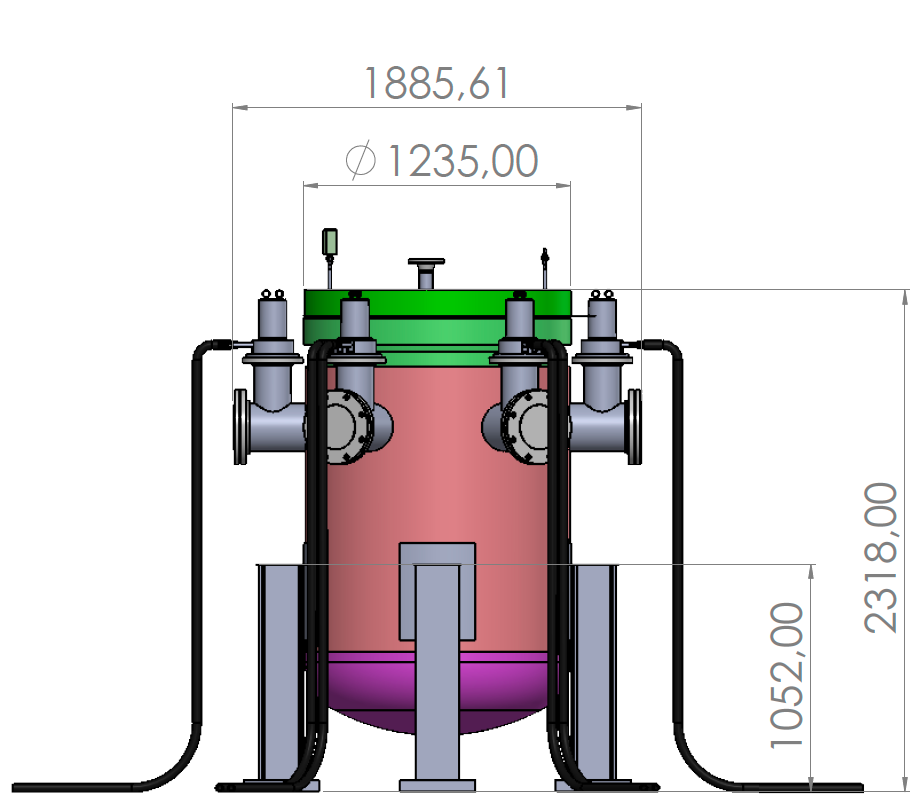} 
    \caption{Left) Design of the Oscura pressure vessel for the operation of the skipper-CCD detector array with 96 SM. Right) Cross section of the Oscura vacuum vessel, with internal lead shield (pink) and external polyethylene shield (blue).
    }
    \label{fig:10kgsketch}
\end{figure}

The power for each readout channel is estimated to be 32~mW as measured in the prototype. This corresponds to less than 1~kW of power for the full system. The current plan is to provide this cooling capacity with closed-cycle cryocoolers~\cite{AL600}. A scheme for the pressure vessel and its radiation shield is shown in Fig.~\ref{fig:10kgsketch}.

\section{Oscura Integration Test}

The Oscura Integration Test (OIT) is a planned engineering test with 10\% of the total mass of the full Oscura detector payload.  The detail configuration of OIT has not been defined. Here, we will consider OIT having a fully instrumented slice, as shown in Figure~\ref{fig:slice}. This slice will have 16 SMs comprising about 2~kg of active CCDs when fully instrumented.

For this engineering test, we do not expect a 100\% production yield, and we assume 1~kg of detectors in good condition.
We do not expect a fully assembled shield for OIT, and we assume 6 inches of lead shielding the detector from environmental radiation. We make no assumptions about an additional neutron shield. For comparison, SENSEI's shielding during its run at the MINOS cavern near detector hall at Fermilab, was only 3 inches of lead~\cite{SENSEI:2020dpa}.

\subsection{Background Assumptions at the MINOS near detector hall}

Operating at MINOS, SENSEI has achieved a background of 3000~DRU~\cite{SENSEI:2020dpa}. 
Here, we will assume that the OIT will achieve a  flat background of 1000~DRU (events~kg$^{-1}$~keV$^{-1}$~day$^{-1}$) at high energies (500~eV to 10~keV).
This means that for a 1~kg detector and 1-year exposure, the number of background events expected in each electron bin is 
\begin{equation}
    N_{bkg}= 1000~\mbox{DRU} \times 0.00375~\mbox{keV} \times 365~\mbox{days} \times 1~\mbox{kg}  \sim 1000 .
\label{eq:bkg}
\end{equation}
where the factor 3.75 was used to convert from eV to electrons~\cite{Rodrigues2021, rodrigues2023unraveling}.
In addition to this contribution from a flat spectrum, there will also be instrumental events, such as thermal dark current. To estimate their contributions, we will consider the rates measured by SENSEI in MINOS~\cite{SENSEI:2020dpa, barak2023sensei}. These rates are summarised in Table~\ref{tab:EventCount}. Since the values of all these rates are well above $N_{bkg}$, we can use them to predict the OIT scientific performance.

\begin{table}[ht]
\begin{center}
\begin{footnotesize}
\caption{Performance of the SENSEI experiment for events containing 1$e^-$ to 6$e^{-}$. 
The efficiency here includes the effect of all selection cuts on the data (see Ref.~\cite{SENSEI:2020dpa} for details). 
The corrected efficiency exposure and the count of observed events after applying the selection criteria are also provided. The last two rows display the projected rates based on the SENSEI dataset. For bins spanning from 3 electrons to 6 electrons, we utilize the 68\% upper limit of 1.14 events to compute the rate.}
\vspace{2mm}
\begin{tabular}{|l|*{12}{c|}}
\hline
Performance / $N_{e}$ & \multicolumn{2}{c|}{1$e^{-}$} & \multicolumn{2}{c|}{2$e^{-}$} & \multicolumn{2}{c|}{3$e^{-}$} & \multicolumn{2}{c|}{4$e^{-}$} & \multicolumn{2}{c|}{5$e^{-}$}    & \multicolumn{2}{c|}{6$e^{-}$} \\ \hline
Efficiency            & \multicolumn{2}{c|}{0.069} & \multicolumn{2}{c|}{0.105} & \multicolumn{2}{c|}{0.325}      & \multicolumn{2}{c|}{0.327} & \multicolumn{2}{c|}{0.331} & \multicolumn{2}{c|}{0.338}   \\ \hline
Exposure~[g-day]          & \multicolumn{2}{c|}{1.38} & \multicolumn{2}{c|}{2.09} & \multicolumn{2}{c|}{9.03}      & \multicolumn{2}{c|}{9.10}     & \multicolumn{2}{c|}{9.23} & \multicolumn{2}{c|}{9.39}  \\ \hline
Observed Events            & \multicolumn{2}{c|}{1312}      & \multicolumn{2}{c|}{5}     & \multicolumn{2}{c|}{0}      & \multicolumn{2}{c|}{0}   & \multicolumn{2}{c|}{0}    & \multicolumn{2}{c|}{0}\\ \hline
Estimated rate [10$^3$ kg-yr]$^{-1}$ & \multicolumn{2}{c|}{347,000} & \multicolumn{2}{c|}{873} 
                                    & \multicolumn{2}{c|}{ $<$ 46.5} & \multicolumn{2}{c|}{ $<$ 46.5}  & \multicolumn{2}{c|}{ $<$ 46.5}& \multicolumn{2}{c|}{ $<$ 46.5}\\ \hline
Estimated rate [kg-day]$^{-1}$ & \multicolumn{2}{c|}{950,000} & \multicolumn{2}{c|}{2,392} 
                                    & \multicolumn{2}{c|}{ $<$ 128} & \multicolumn{2}{c|}{ $<$ 128} 
                                    & \multicolumn{2}{c|}{ $<$ 128} & \multicolumn{2}{c|}{ $<$ 128}
                                    \\ \hline
\end{tabular}
\end{footnotesize}
\label{tab:EventCount}
\end{center}
\end{table}%

\section{Search for Millicharged particles with the NuMI beam}

\begin{figure}[ht]
    \centering
    \includegraphics[width=0.6\textwidth]{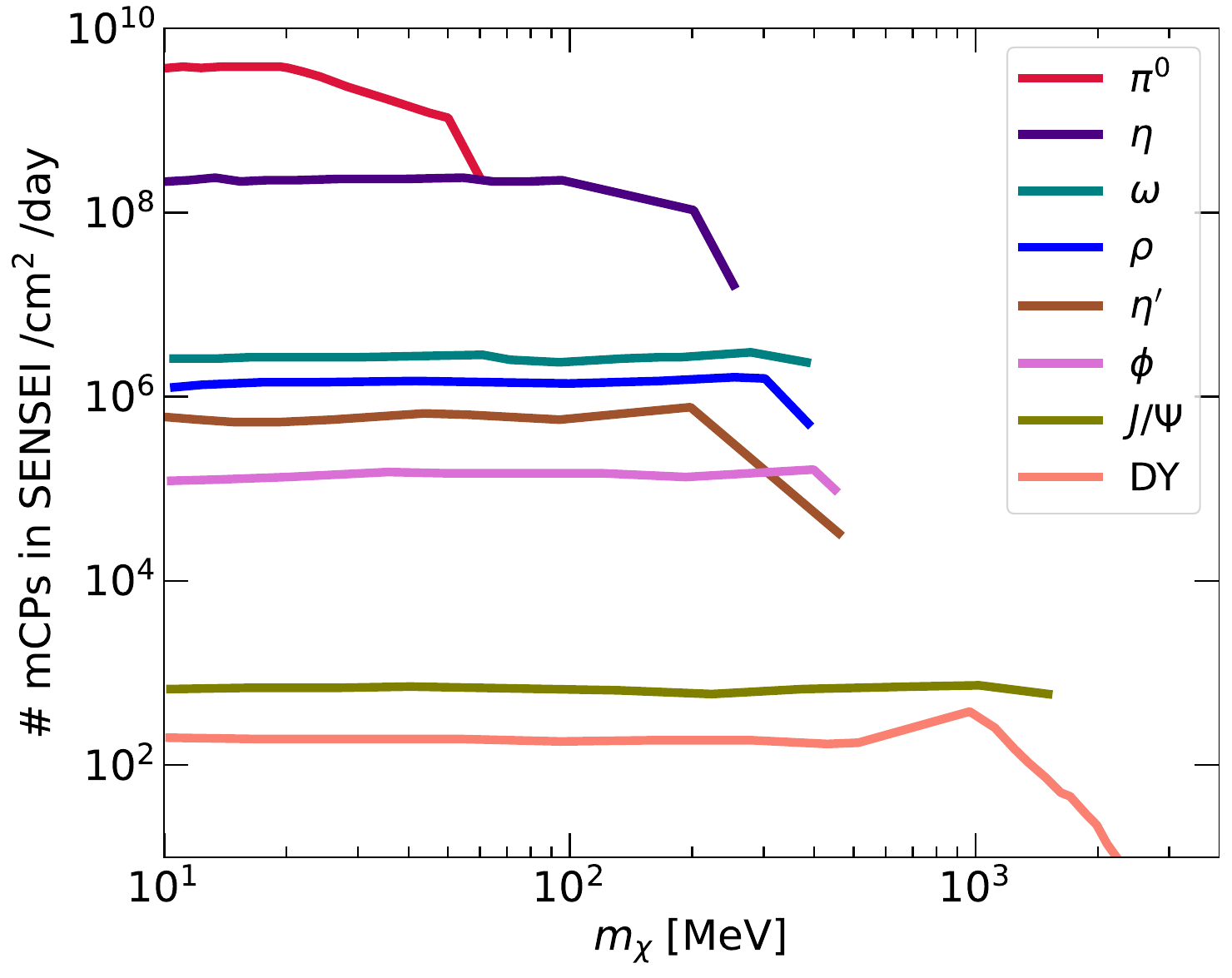} 
    \caption{Conservative numbers of mCPs per cm$^{-2}$ per day $\sim$ 500~m from the target. Fluxes were obtained with $\varepsilon=1$ and 2 $\times 10^{18}$ Protons on Target, integrating over the mCP energies. Flux scales proportionally to $\varepsilon^2$. Adapted from \cite{mcpRoni2019}.}
    \label{fig:mcpflux}
\end{figure}

There exists a unique opportunity for the OIT to be installed in the MINOS near-detector hall to look for millicharged particles (mCP) produced by the NuMI beam. A first analysis of this search was conducted by the Fermilab ArgoNeuT Experiment~\cite{mcpRoni2019, ArgoNeuT:2019ckq}, and recently, SENSEI at MINOS has demonstrated the capability of the Skipper-CCD technology by establishing word leading limits on mCP parameters in the MeV mass range.~\cite{barak2023sensei}. Accelerator neutrino beams are produced by a proton colliding with a fixed target, in the particular case of the NuMI beamline at Fermilab, protons are accelerated with an energy of 120~GeV and collide with a fixed graphite target. Neutrinos are produced mainly by charged pion decays, but there are other types of particles coming from interactions between the protons and the target. mCPs could be produced through photon-mediated processes from decays of scalar, pseudoscalar, and vector mesons, or direct Drell-Yan productions.
The production fluxes, derived from 2 $\times 10^{18}$ Protons on Target (POT), are illustrated in Figure~\ref{fig:mcpflux}. Here, the charge $\varepsilon$ of the mCP was assumed to be 1. It is important to note that the flux scales proportionally to $\varepsilon^2$. The MINOS near-detector hall, positioned approximately 500~m away from the target hall, presents an ideal location to place an experiment looking for these kind of particles. This hall could potentially house the OIT setup, allowing it to be exposed to the incoming flux of mCPs.

\begin{figure}[ht]
    \centering
    \includegraphics[width=.35\textwidth]{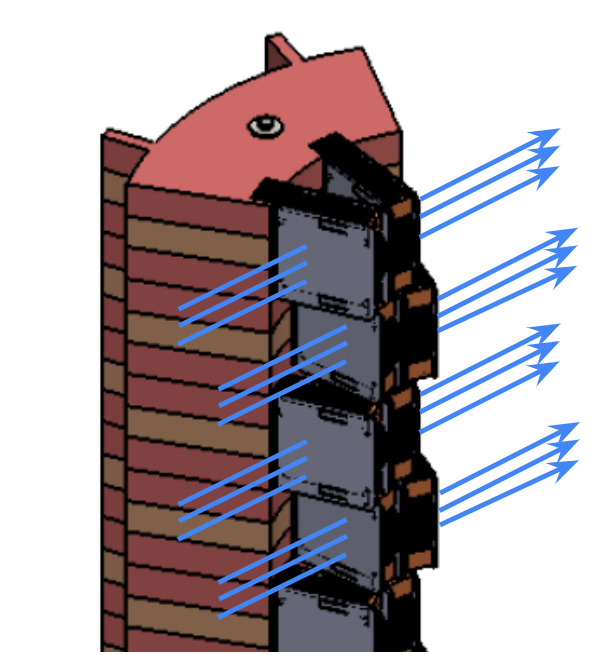} 
    \caption{mCP tracks passing through the Oscura Integration Test detector.}
    \label{fig:mcptrack}
\end{figure}

Due to the mCPs produced in the NuMI beamline being highly boosted, angular deflections with the matter between the target and the detector can be disregarded. This makes the expected signal collinear with the proton beam and have a uniform flux on the entire OIT detector. The Oscura slice in OIT will be vertically placed, and each mCP will traverse two Oscura supermodules as shown in Fig.~\ref{fig:mcptrack}. In this configuration, OIT will function as an mCP tracker with $N$=32 silicon tracking layers. 

\subsection{Mean free path and intrinsic efficiency} 

The mean free path $\lambda$ for particles passing the detector as a function of the cross-section can be expressed as
\begin{equation}
\lambda = \Bigl[n_{det} \sigma(E_{r}^{min} ,E_{r}^{max})\Bigr]^{-1},
\label{eqn:lambda}
\end{equation}

\noindent where $n_{det}$ is the electron number density of the detector material and $\sigma(E_{r}^{min},E_{r}^{max})$ is the scattering cross-section. For further discussion, it is useful to highlight the dependence of $\lambda$ on the quantities of interest. Hence, we can write the following simplified relation when the recoil energy $E_{recoil}$ is above 20~eV,

\begin{equation}
    \lambda \propto \frac{E_{recoil}}{\varepsilon^2}.
    \label{eqn:lambda_dependence}
\end{equation}

In regards to cross-section calculations, mCPs are expected to be ultra-relativistic charged particles that arrive at the detector with a velocity parameterised by $\beta=p/E$. Given that these particles have a charge of $q=\varepsilon e$, where $\varepsilon$ can be a real number between 0 and 1 and $e$ is the charge of the electron, mCPs are expected to interact electromagnetically within our detector. Following \cite{Plasmon:forthcoming},  the interaction cross section between the mCPs and silicon can be described by,
 
\begin{equation}    
    \label{energy-loss-fermi}
     \frac{d\sigma}{d\omega} = \frac{8\alpha\varepsilon^2}{n_e \beta^2}\int_0^\infty dk \bigg\{\frac{1}{k}\mathrm{Im}\left(-\frac{1}{\epsilon(\omega, k)}\right)+k\left(\beta^2 - \frac{\omega^2}{k^2}\right)\mathrm{Im}\left(\frac{1}{-k^2 + \epsilon(\omega, k)\omega^2}\right)\bigg\},
\end{equation}
where $\alpha$ is the fine structure constant,  $k_\mu = (\omega, k)$, $n_e$ the number of electrons per unit volume, and $k=|k|$. 

\begin{figure}[ht]
    \centering    \includegraphics[width=.6\textwidth]{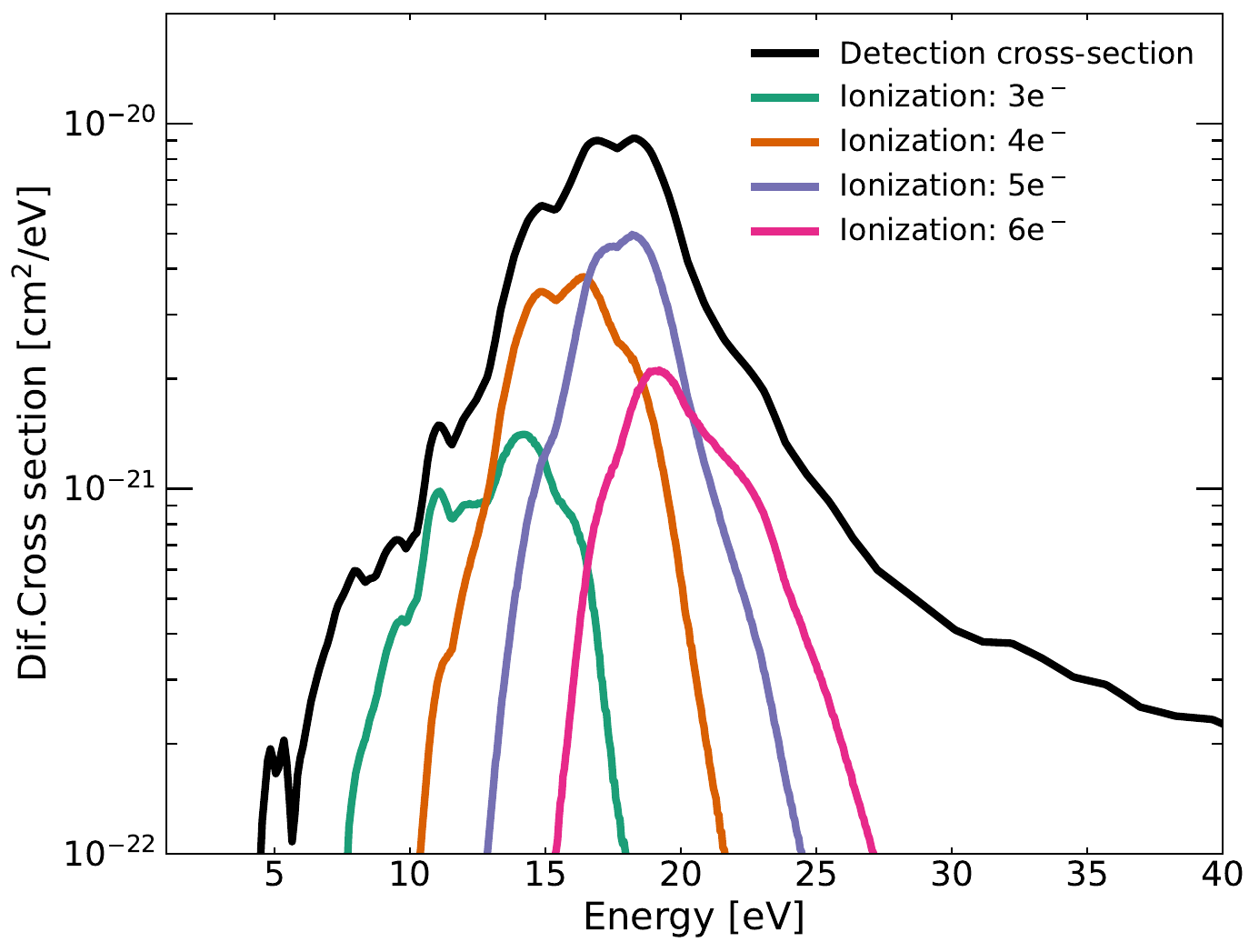}
    \caption{Interaction cross section convolved with the ionization yield for 3$e^-$, 4$e^-$, 5$e^-$ and 6$e^-$ signals. The cross-section was calculated using $\varepsilon$=1.}
    \label{fig:sigma_effi}
\end{figure}

The expression for $\sigma(E_{r}^{min},E_{r}^{max})$ can be derived by integrating the differential cross-section between the minimum recoil energy $E_{r}^{min}$ and the maximum recoil energy allowed for the electrons $E_{r}^{max}$. For a more precise calculation, the ionization yield~\cite{Ramanathan2020} for signals between $3e^-$ and $6e^-$ has been convolved with $d\sigma/dE_{r}$ to account for the probability of an electron recoil to generate a given number of electron hole pairs. The results for the cases relevant to this work are shown in Fig.~\ref{fig:sigma_effi}. 
Taking into consideration that the range between 3 to 6 electrons holds the highest cross-section, we will restrict our analysis to this interval and interactions producing electrons within this range, henceforth we will simply refer to signal form this channel as an events.

\subsection{Expected number of events and geometry dependency}

This work will concentrate on two different approaches to detect mCPs within the $N$=32-layer silicon tracker: specifically, analysing single events as described in Ref.~\cite{barak2023sensei}, and examining tracks involving at least two events aligned with the NuMI beam target following the ideas discussed in Ref.~\cite{mcpRoni2019}.

We will start assuming that $\varepsilon$ is sufficiently small for $\lambda(\varepsilon)$ to greatly exceed the Skipper-CCD thickness $\Delta z = 725 \, \mu \text{m}$. As a reasonable approximation, the probability for a mCP to produce an event in a given layer is $p = \Delta z / \lambda$.
Then, for a tracker with $N$ layers, the expected number of events produced by a single mCP is 
\begin{equation*}
\mu= N p = L/ \lambda
\end{equation*}
where $L=N \Delta z$ represents the effective thickness of the full tracker.
Hence, the probability of a mCP producing at least $k$ events is determined using the binomial distribution:
\begin{equation*}
 \xi(k|N,p) = 1-\sum_{m=0}^{k-1} \binom Nm p^m (1-p)^{N-m}.
 \label{eq:binomial}
\end{equation*}
that under the Poissonian approximation, valid when $p \rightarrow 0$ and $N \gg 1$, becomes:
\begin{equation}
 \xi(k|\mu) = 1-\sum_{m=0}^{k-1} \mu^m e^{-\mu}/m!.
 \label{eq:poisson}
\end{equation}

\begin{figure}[ht]
    \centering
    \includegraphics[width=0.6\textwidth]{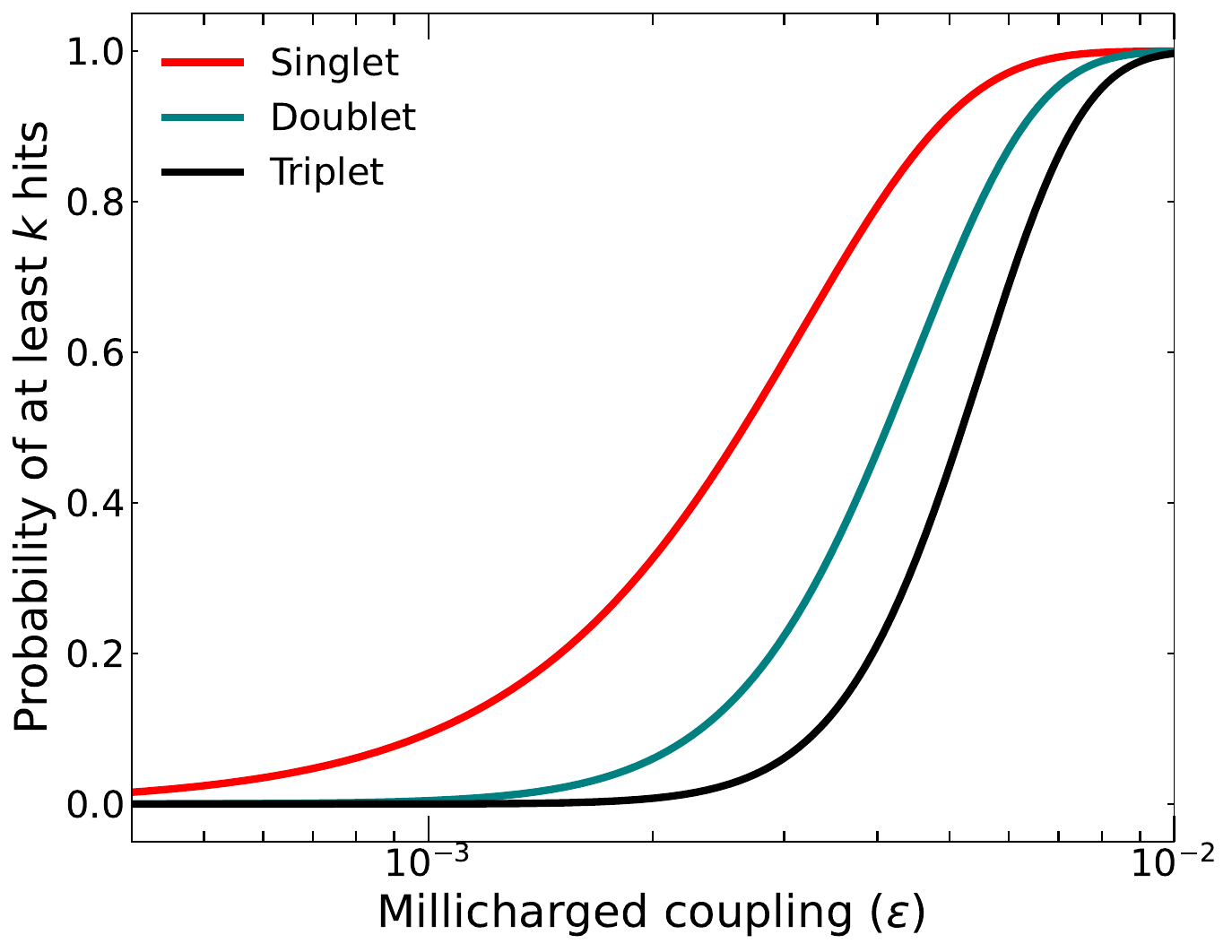} 
    \caption{Probability for singlet, doublet, and triplet events.}
    \label{fig:proba_accum}
\end{figure}

From this point forward, we will denote a minimum of one single event as \textit{singlet} ($k$=1), a minimum of two events as a \textit{doublet} ($k$=2), and a minimum of three events as a \textit{triplet} ($k$=3).
Their probabilities of occurrence are plotted as a function of $\varepsilon$ in Fig.~\ref{fig:proba_accum}. Observe that the dependence of $\mu$ on $\varepsilon$ occurs via $\lambda$.

In the pursuit of determining the expected number of events $\langle n \rangle$ detected by the OIT, resulting from mCPs generated at the NuMI beam, Eq.~\eqref{eq:poisson} has to be convolved with the flux  $\phi(\varepsilon^2, m_{\chi})$ of mCPs (number of particles of mass $m_{\chi}$ per cm$^2$ per day), that is:

\begin{equation}
    \langle n(\varepsilon,m_{\chi}) \rangle = A \, E \, \int \phi(\varepsilon^2, E_{\chi}, m_{\chi}) \, \xi(\varepsilon, E_{\chi}, m_{\chi}) dE_{\chi},
\end{equation}
where the scaling factors are the area of the detector which is facing the beam $A$ = $N_{pix} \times \Delta x \times \Delta y = N_{pix} \times 15 \, \mu \text{m} \times 15 \, \mu \text{m}$, and the exposure $E$.
Although the full slice will have a surface area consisting of 8 MCMs (8 × 16 sensors of 1.35 Mpix each), we have assumed a 50\% yield for this test (given 1~kg instead of 2~kg). To account for this inefficiency, we will use a total of $N_{pix}= 86.4 \times 10^6$ pixels (half of the total).

Taking into account that in the ultra-relativistic regime, the cross-section dependency on the mCP energy $E_{\chi}$ is negligible  and the flux in Fig. \ref{fig:mcpflux} has been already integrated over $E_{\chi}$, it is an excellent approximation to write:
 
\begin{equation}
\langle n(\varepsilon,m_{\chi}) \rangle = A \, E\, \phi(\varepsilon^2,m_{\chi}) \, \xi(\varepsilon,m_{\chi}).
\label{eq:expected_events}
\end{equation}

It is worth noting that in the limit of $\mu \ll 1$, which holds when $\lambda \gg L$, Eq.~\eqref{eq:expected_events} becomes
\begin{equation*}
 \langle n(\varepsilon, m_{\chi}, L) \rangle \propto A \, E \, \phi(\varepsilon^2,m_{\chi}) \, (L\, \varepsilon^{2})^n = V \, E \, \phi\prime(m_{\chi}) L^{n-1} \, \varepsilon^{2(n+1)},
 \label{eq:approx2}
\end{equation*}
that for the case of doublets, result to be:
\begin{equation*}
 \langle n(\varepsilon, m_{\chi}, L) \rangle \propto V \, E \, L \, \phi\prime(m_{\chi}) \, \varepsilon^6.
 \label{eq:approx3}
\end{equation*}
Thus we find that for a given flux of mCPs, exposure, and detector volume $V$, the expected number of doublets increases linearly with the length of the detector and scales with $\varepsilon^6$.

\subsection{Background estimation}

As mentioned above, the search for mCPs could be performed by looking for single-scattering events or multiple-scatterings in the silicon tracker. Random coincidences of non-correlated events would produce fake tracks, whose rate depends on the geometry of the detector and its time resolution. 
For OIT, we estimate the experimental background based on the performance observed for SENSEI in Table~\ref{tab:EventCount}, assuming that the detector is fully read out in 1 day (readout time 24 hrs).  Any pair of events aligned with the beam during each 1-day exposure will constitute a background track for the mCP search.

The number of random coincidences producing tracks with at least $b$ events for a tracker with $N_{pix}$ pixels on each layer is 
\begin{equation}
N_{bkg}^{tracks} = N_{pix} \times \xi(b|N,p_{bkg}), \label{eq:faketracks}
\end{equation}
where $p_{bkg}$ represents the probability of a single-pixel having between 3 to 6 electrons within a day, generated with the background rates outlined in Table \ref{tab:EventCount}. 
It can be calculated as the ratio between background events per kg and the total number of pixels in the same mass, that is $p_{bkg} = (4 \times 128)/(N \times N_{pix})$. Edge effects and uncertainties in the position of the event were not taken into account.
As a result, while an estimated of 512 single events per kilogram per day are expected (obtained by summing the values in the last four columns of Table \ref{tab:EventCount}), we anticipate approximately $1.5 \times 10^{-3}$ tracks involving a minimum of two events aligned with the NuMI beam target, originating from background sources. Furthermore, the projection extends to approximately $1 \times 10^{-8}$ tracks involving at least three events under the same conditions. The last two constitute the background for doublets and triplets searches and essentially means that looking for tracks results in a background-free strategy.

\subsection{Sensitivity}

To consider the effect of backgrounds, we use the mCP flux from the beam in Fig.~\ref{fig:mcpflux} and the interaction cross-section to estimate the experiment's sensitivity for detecting those mCPs with a one-year exposure and non-zero background.

The analysis can be done using $n_{signal}$, obtained by subtracting the number of events observed when the beam is OFF ($n_{bkg}$) from the number of events observed when the beam is ON ($n_{ON}$).

Calculating the uncertainty in $n_{signal}$ as $\sigma_{signal}^2=\sigma_{ON}^2+\sigma_{bkg}^2$, a simple method to determine a 90\% confidence level in the presence of background is to demand the signal to be 1.28~$\times~\sigma_{signal}$ above zero. 
Assuming Poissonian statistics for the number of observed events i.e.
$\sigma_{ON}^2~\sim~n_{ON}$, and a similar exposure with beam ON and OFF, we have $\sigma_{bkg}^2~\sim~\sigma_{ON}^2$. 
Taking into consideration all of this, the upper limit can be obtained by asking for $n_{ON}-n_{bkg}\geq1.28~\times~\sqrt{2 N_{bkg}^{tracks}}$.
To generalize for different exposure times, a likelihood ratio analysis can be applied as proposed in reference~\cite{Cowan2013}.

\begin{figure}[t]
    \centering
    \includegraphics[width=0.75\textwidth]{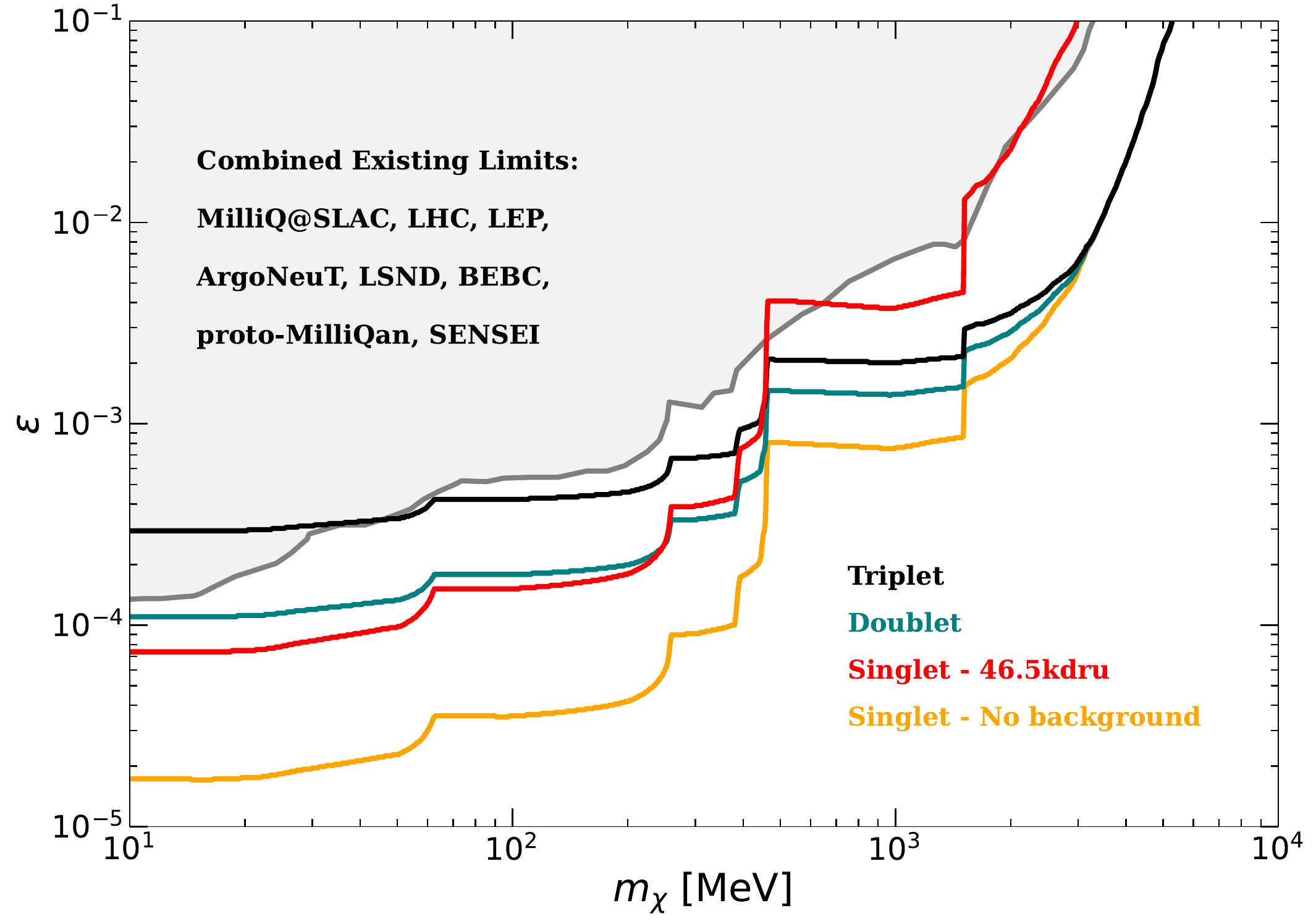}
    \caption{Exclusion limits with a 90\% confidence level for the different strategies searching for singlets (yellow and red), doublets (blue) and triplets (black) considering a 1~year and 1~kg experiment.}
    \label{fig:limits}
\end{figure}

Given the mCP flux from the beam in Fig.~\ref{fig:mcpflux} and the interaction cross-section from Eq.~\eqref{energy-loss-fermi}, we estimate the sensitivity of OIT for the detection of mCPs. 
We assume 1~kg of active mass and 2 $\times 10^{18}$ POT. The limits produced by each strategy described here are shown in Fig.~\ref{fig:limits}. 
The OIT forecast is compared with the current limits of other mCP detection experiments at accelerators~\cite{snowmass2021_beamdark}.

\section{Discussion}

Although the background-free singlet search provides the best sensitivity in most of the mass range, including background brings the constraint closer to that of the doublet search strategy.
For masses below 220 MeV, the singlet strategy provides the strongest constraints and thereafter, the doublets strategy becomes more advantageous due to its resilience to the background. 
The triplet strategy does not outperform the doublet strategy, and it only becomes advantageous in higher background scenarios, as will be discussed in the next section.
It is worth mentioning that the expected limit using the doublet strategy below 220 MeV is better than projections of any other mCP detection experiments, such as FORMOSA~\cite{Foroughi-Abari:2020qar} and Fermini~\cite{fermini}.

\begin{figure}[t]
    \centering
    \includegraphics[width=0.65\textwidth]{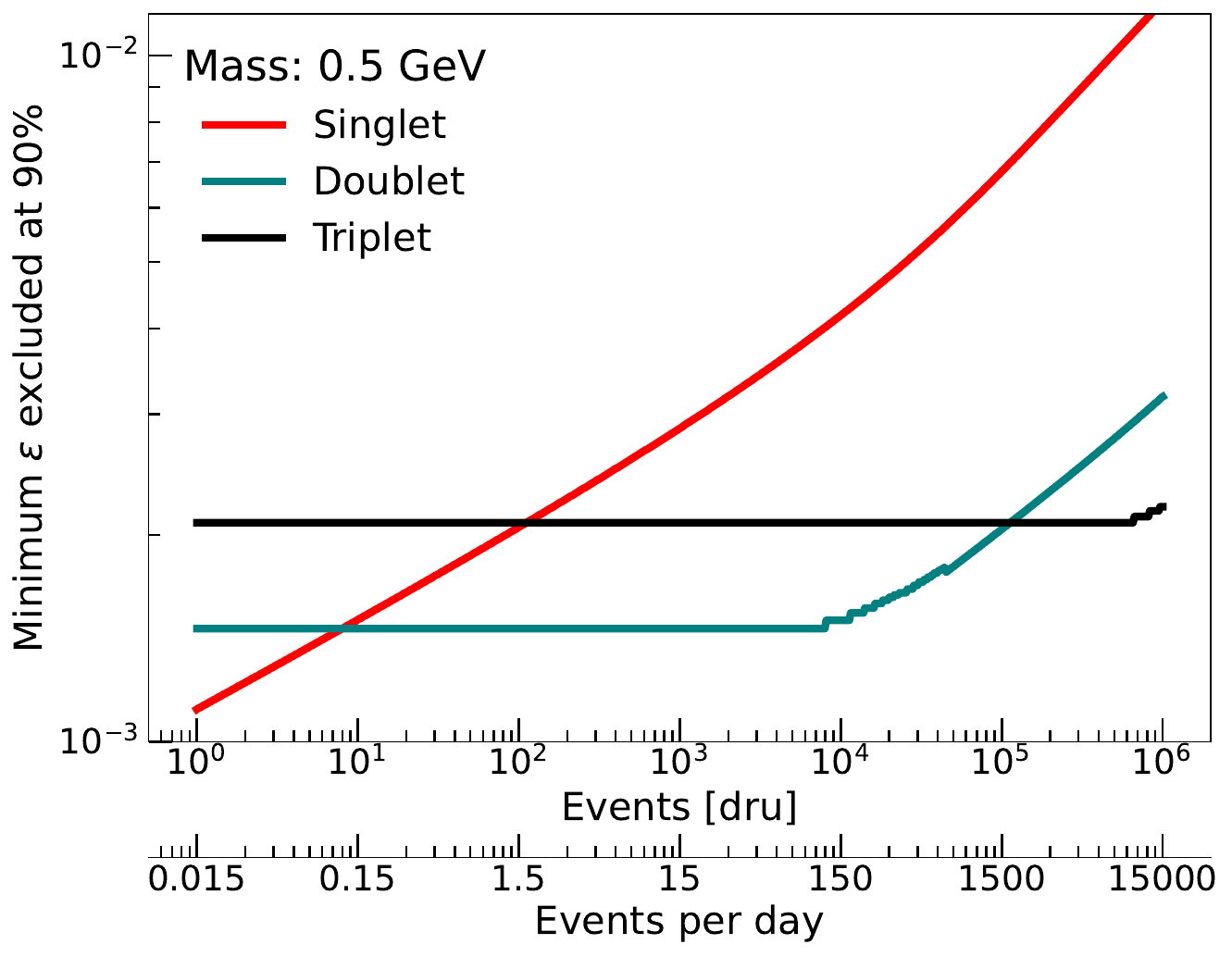}
    \caption{Dependence of the minimum $\varepsilon$ with the background rate for a mCP mass of $0.5$~GeV. The singlet search strategy shows a strong dependence on the background rate while the other two probe to be more robust.}
    \label{fig:limits2}
\end{figure}

The optimal approach hinges on the detector's geometry and background rate. To discern the circumstances favoring a request for triplets over doublets, an exploration of elevated background scenarios becomes necessary. 
Fig.\ref{fig:limits2} illustrates a comparison among the three strategies for a mCP mass of 0.5 GeV. The doublet search strategy is the preferred choice until the background reaches 100 kdru, at which point searching for triplets provides better exclusion limits. This is an anticipated outcome as the superiority of the triplet strategy stems from its greater resilience against the background.

As can be seen in Fig.~\ref{fig:proba_accum}, above $\varepsilon \sim 10^{-2}$, the probability of obtaining singlets, doublets, or triplets is essentially 1, therefore for higher masses, we are flux-limited (see Fig.~\ref{fig:mcpflux}).  On the other hand, for lower masses, which implies lower values of $\varepsilon$ and higher fluxes, the background does not play a crucial role, and the singlet strategy is preferable.

In summary, we explored the science potential for the Oscura Integration Test (OIT). This test is planned as part of the development of the Oscura 10~kg Skipper-CCD experiment~\cite{Oscura2022} and corresponds to approximately $10\%$ of the total detector payload. We assume that OIT is installed in the MINOS near-detector hall with a modest shield (6-inch lead), comparable to the configuration used by SENSEI in its MINOS run~\cite{SENSEI:2020dpa, barak2023sensei}.
This analysis shows that OIT would provide the opportunity for a vigorous search for low-mass mCPs, covering an unexplored region of the parameter space, improving the current mCP limits and, at low masses, it would reach regions of $\varepsilon$ not accessible by other proposed experiments.

\bibliographystyle{JHEP}
\bibliography{biblio.bib}

\end{document}